\documentclass{llncs}

\newbox\tempa
\newbox\tempb
\newdimen\tempc
\def\mud#1{\hfil $\displaystyle{\mathstrut #1}$\hfil}
\def\rig#1{\hfil $\displaystyle{#1}$}
\def\irulehelp#1#2#3{\setbox\tempa=\hbox{$\displaystyle{\mathstrut #2}$}%
                        \setbox\tempb=\vbox{\halign{##\cr
        \mud{#1}\cr
        \noalign{\vskip\the\lineskip}
        \noalign{\hrule height 0pt}
        \rig{\vbox to 0pt{\vss\hbox to 0pt{${\; #3}$\hss}\vss}}\cr
        \noalign{\hrule}
        \noalign{\vskip\the\lineskip}
        \mud{\copy\tempa}\cr}}
                      \tempc=\wd\tempb
                      \advance\tempc by \wd\tempa
                      \divide\tempc by 2 }
\def\irule#1#2#3{{\irulehelp{#1}{#2}{#3}
                     \hbox to \wd\tempa{\hss \box\tempb \hss}}}
\def\birulehelp#1#2#3{\setbox\tempa=\hbox{$\displaystyle{\mathstrut #2}$}%
                        \setbox\tempb=\vbox{\halign{##\cr
        \mud{#1}\cr
        \noalign{\vskip\the\lineskip}a
        \noalign{\hrule height 0pt}
        \phantom{$#3$}
                \rig{\vbox to 0pt{\vss\hbox to 0pt{${\; #3}$\hss}\vss}}\cr
        \noalign{\hrule}
        \noalign{\vskip\the\lineskip}
        \mud{\copy\tempa}\cr}}
                      \tempc=\wd\tempb
                      \advance\tempc by \wd\tempa
                      \divide\tempc by 2 }
\def\birule#1#2#3{{\birulehelp{#1}{#2}{#3}
                     \hbox to \wd\tempa{\hss \box\tempb \hss}}\phantom{#3}}

\begin{document}
\title{A constructive proof of Skolem theorem for constructive logic}
\author{Gilles Dowek \and Benjamin Werner}  
\institute{Projet LogiCal\\ P{\^o}le Commun de Recherche en
  Informatique du Plateau de Saclay \\
\'Ecole polytechnique, INRIA, CNRS
  and Universit\'e de Paris-Sud\\
  LIX, \'Ecole polytechnique, 91128
  Palaiseau Cedex, France\\
{\tt\{Gilles.Dowek,Benjamin.Werner\}@polytechnique.fr}}
\maketitle

\begin{abstract}
If the sequent $\Gamma \vdash \forall \overline{x} \exists y~A$ is
provable in first order constructive natural deduction, then the
theory $\Gamma, \forall \overline{x}~(f(\overline{x})/y)A$, where $f$
is a new function symbol, is a conservative extension of $\Gamma$.
\end{abstract}

Skolem theorem asserts that if the sequent $\Gamma \vdash \forall
\overline{x} \exists y~A$ is provable in first order constructive
logic, then the theory $\Gamma, \forall
\overline{x}~(f(\overline{x})/y)A$, where $f$ is a new function
symbol, is a conservative extension of $\Gamma$. That is if the
sequent $\Gamma, \forall \overline{x}~(f(\overline{x})/y)A \vdash B$
is provable and $f$ does not occur in $B$, then the sequent $\Gamma
\vdash B$ is also provable.  We give in this note a constructive
proof, providing an algorithm transforming proofs of one sequent into
proofs of the other. Our proof follows the lines of \cite{Maehara}
with two differences. First, our logic is first-order constructive
logic and not classical, second we use a formulation of the deduction
rules based on natural deduction and not on sequent calculus.

We shall use two results about constructive natural deduction. 
First, the cut elimination theorem: when a sequent has a proof it
also has a 
proof free of cuts, including the so-called commutative cuts. Second,
the subformula property: a cut free proof of a sequent contains only
subformulas of this sequent.  
Recall that the subformulas of a formula $B$ are defined as follows.
If $B$ is atomic, then its only subformula is $B$ itself.
If $B$ is $C \Rightarrow D$, $C \wedge D$ or $C \vee D$, then 
the subformulas of $B$ are $B$ and the subformulas of $C$ and $D$.
It $B$ is $\forall x~C$ or $\exists x~C$, then the subformulas of $B$ are 
$B$ and the subformulas of all the propositions of the form 
$(t/x)C$ for some term $t$. 
The rules of constructive natural
deduction and proofs of both results can be found, for instance, in
\cite{Prawitz}.

We shall also use the admissibility of the weakening rule, i.e. the
fact that if the sequent $\Gamma \vdash C$ is provable, then so is the
sequent $\Gamma, B \vdash C$. This lemma can be proved by a simple
induction on proof structure.

Let $\overline{x}=x_1,\dots,x_n$ be a finite sequence of distinct
variables and $C$ a proposition, we write $\forall\overline{x}~C$ for
$\forall x_1\dots\forall x_n~C$. Similarly, if
$\overline{t}=t_1,\dots,t_n$ is a finite sequence of terms and $f$ a
function symbol of arity $n$, we write $f(\overline{t})$ for the term
$f(t_1,\dots,t_n)$. Finally, we write $\overline{t}/\overline{x}$ for
the simultaneous substitution $t_1 / x_1, \dots, t_n / x_n$.

Throughout this note, we consider a fixed proposition $A$ where the
symbol $f$ does not occur and whose free variables are among $x_{1},
\dots, x_{n}, y$. A {\em partial instance} is a proposition of the
form $(t_1/x_1,\dots,t_i/x_i)\forall x_{i+1} \dots \forall x_n~ 
(f(\overline{x})/y)A$ with $0 \leq i < n$.  A {\em total instance} is
a proposition of the form
$(\overline{t}/\overline{x})(f(\overline{x})/y)A$.

A {\em $f$-term} is a term of the form $f(\overline{t})$.  The
occurrence of a subterm $u$ is said to be {\em frozen} in a
proposition $C$ if all variable occurrences of $u$ are free in $C$.
As a consequence of the subformula property, if in a sequent $\Gamma
\vdash B$ all $f$-terms are frozen, then all $f$-terms are also frozen
in a cut free proof of this sequent.
Another consequence of the subformula property is that if all
$f$-terms are frozen in $\Gamma$ and $B$, then in a cut free proof of
the sequent $\Gamma, \forall \overline{x}~(f(\overline{x})/y)A \vdash
B$, any proposition $C$ is either a partial instance or all $f$-terms
are frozen in $C$.

\begin{proposition}[Proof pruning]
\label{pruning}
Let $\Gamma$ be a context and $B$ be a proposition where all $f$-terms
are frozen.  If the sequent $\Gamma \vdash B$ is provable, then so is 
the sequent $\sigma \Gamma \vdash \sigma B$, where $\sigma$ is the 
function 
mapping terms to terms and propositions to propositions
replacing a $f$-term $f(\overline{u})$ by a variable $z$ 
not occurring in $\Gamma$ and $B$.
\end{proposition}

\proof{The sequent $\Gamma \vdash B$ is provable, thus it has a cut
free proof and all the $f$-terms are frozen in this proof. Without
loss 
of generality, we can assume that the variable $z$ does not appear in
this proof. By induction on the structure of this proof, 
we build a proof of the sequent $\sigma \Gamma \vdash \sigma B$ 
using, for quantifier rules, the fact that $\sigma
((t/x)P) = (\sigma t/x) \sigma P$ when $x$ occurs in no $f$-term
of $P$.}

\begin{proposition}[Elimination of hypotheses]
\label{erase}
Let $\Gamma$ be a context and $B$ be a proposition where all $f$-terms
are frozen.  Let $\overline{u}$ be a term sequence such that the term
$f(\overline{u})$ does not occur in $\Gamma$
and $B$.  If the sequents
$\Gamma \vdash \forall \overline{x}\exists y~A$ and $\Gamma,
(\overline{u}/\overline{x}, f(\overline{u})/y)A \vdash B$ are
provable then so is the sequent $\Gamma \vdash B$.
\end{proposition}

\proof{
Let $z$ be a fresh variable. 
Let $\sigma$ be
the function 
replacing the subterm $f(\overline{u})$ by the
variable $z$. 
By Proposition \ref{pruning}, the
sequent 
$\Gamma, (\overline{u}/\overline{x}, z/y)A \vdash B$ 
is provable.
From the proofs of
the sequent $\Gamma \vdash \forall \overline{x}\exists y~A$ and
$\Gamma, (\overline{u}/\overline{x}, z/y)A \vdash B$ we build a proof of
the sequent $\Gamma \vdash B$.}

\begin{proposition}[Partial instances]
\label{partial}
Let $\Gamma$ be a context and $B$ be a proposition where all $f$-terms
are frozen.  Let $\pi$ be a cut free proof of the sequent $\Gamma,
\forall \overline{x}~(f(\overline{x})/y)A \vdash B$.  If the proof
$\pi$ contains an occurrence of a sequent $\Delta \vdash C$, where $C$
is a partial instance of $\forall \overline{x}~(f(\overline{x})/y)A$,
then this sequent is the premise of an elimination of the
universal quantifier.
\end{proposition}

\proof{By induction on the depth of the occurrence of the sequent $\Delta
\vdash C$ in the proof. This sequent is not
the conclusion of the proof as all $f$-terms are frozen in $B$, thus 
it is the premise of some rule. 

If this rule is an introduction rule then the conclusion is a
proposition where $f$ 
occurs with a bound variable as one of its arguments, thus
it is a partial instance
of $\forall\overline{x}~(f(\overline{x})/y)A$. 
Applying the induction hypothesis, the next rule is an elimination of
the universal quantifier,
contradicting the fact that the proof is cut free.

The sequent $\Delta \vdash C$ cannot be the minor premise of a
elimination of the implication because the major premise would
then be a proposition that is not a partial instance but where $f$
occurs with a bound variable as one of its arguments.

If it is a minor premise of an elimination of the disjunction or
of the existential quantifier then the conclusion of this elimination
rule is of the form $\Delta' \vdash C$. Applying the induction
hypothesis, the next rule is $\forall$-elim, contradicting the fact
that the proof is free of commutative cuts.

Thus this sequent is the major premise of an elimination rule and this
rule is the elimination of the universal quantifier.}

\begin{theorem}
Let $\Gamma$ be a context and $B$ be a proposition where the symbol
$f$ does not occur. If the sequents $\Gamma \vdash \forall
\overline{x} \exists y~A$ and $\Gamma, \forall
\overline{x}~(f(\overline{x})/y)A \vdash B$ are provable, then the
sequent $\Gamma \vdash B$ is provable.
\end{theorem}

\proof{We prove, more generally, that if $\Gamma$ is a context and $B$
is a proposition where all $f$-terms are frozen and
$f(\overline{u}_1), \dots, f(\overline{u}_k)$ are the $f$-terms
occurring in $\Gamma$ and $B$, then if the sequents $\Gamma \vdash
\forall \overline{x}\exists y~A$ and $\Gamma, \forall
\overline{x}~(f(\overline{x})/y)A \vdash B$ are provable then the
sequent $\Gamma, \Delta \vdash B$ where $\Delta =
(\overline{u}_1/\overline{x}, f(\overline{u}_1)/y)A, \dots,
(\overline{u}_k/\overline{x}, f(\overline{u}_k)/y)A$ is provable.

The proof proceeds by induction on the structure of a cut free proof
of $\Gamma, \forall \overline{x}~(f(\overline{x})/y)A \vdash B$.  The
last rules derives $\Gamma, \forall \overline{x}~(f(\overline{x})/y)A
\vdash B$ from premises $\Gamma_1, \forall
\overline{x}~(f(\overline{x})/y)A \vdash B_1$, \dots, $\Gamma_p,
\forall \overline{x}~(f(\overline{x})/y)A \vdash B_p$.

\begin{itemize}
\item If one of the propositions $B_1, \dots, B_p$ is a partial
instance of $\forall \overline{x}~(f(\overline{x})/y)A$, then by
Proposition \ref{partial} the last rule is an elimination of the
universal quantifier, the proposition $B$ has the form
$(\overline{t}/\overline{x}, f(\overline{t})/y)A$ and $\Delta$ 
contains $B$. Thus the sequent
$\Gamma, \Delta \vdash B$ is provable with the axiom rule.

\item Otherwise, all the $f$-terms are frozen in all the $B_i$'s.
We check, by a case analysis on the rule used, that all the 
$f$-terms are frozen in all the $\Gamma_i$'s.  
By
induction hypothesis, the sequents $\Gamma_1, \Delta_1 \vdash
B_1$, \dots, $\Gamma_p, \Delta_p \vdash B_p$ are provable where
$\Delta_{i}$ is the set of total instances corresponding to
the frozen $f$-terms of $\Gamma_i$ and $B_i$. Let $\Delta'$ be
the union of all these sequences. By weakening, the sequents
$\Gamma_1, \Delta' \vdash B_1$, \dots, $\Gamma_p, \Delta'
\vdash B_p$ are provable.

We now check that the sequent $\Gamma, \Delta' \vdash B$ can be
proved from $\Gamma_1, \Delta' \vdash B_1$, \dots, $\Gamma_p,
\Delta' \vdash B_p$. The only point to check is that the
eigenvariable conditions are verified in the case of the introduction
rule of the universal quantifier and the elimination of the
existential quantifier. In the first case, the proof has the form
$$
\irule{\irule{\dots}
             {\Gamma_1, \forall \overline{x}~(f(\overline{x})/y)A
             \vdash B_1}
             {}
     }
     {\Gamma_1, \forall \overline{x}~(f(\overline{x})/y)A \vdash \forall z~B_1}
     {}$$
and the variable $z$ is not free in $\Gamma_1, \forall
\overline{x}~(f(\overline{x})/y)A$. 
Notice that, by hypothesis, the proposition $\forall z~B_1$ is 
not a partial instance. 
Hence, the variable $z$ does not occur in an argument of $f$ in
$B_1$. As moreover $z$ is not free in $\Gamma_1$, it is not free in $\Delta_1$
and from $\Gamma_1, \Delta_1
\vdash B_1$, we can deduce $\Gamma_1, \Delta_1 \vdash \forall
z~B_1$.

In the second case, the proof has the form
$$\irule{\irule{\dots}
               {\Gamma_1, \forall \overline{x}~(f(\overline{x})/y)A
                                                       \vdash \exists z~B_1}
               {}
          ~~~
         \irule{\dots}
               {\Gamma_1, \forall \overline{x}~(f(\overline{x})/y)A,
                                                              B_1 \vdash B_2}
               {}
        }
        {\Gamma_1, \forall \overline{x}~(f(\overline{x})/y)A  \vdash B_2}
        {}$$
and the variable $z$ is not free in $\Gamma_1, \forall
\overline{x}~(f(\overline{x})/y)A$ and $B_2$.  Notice that the
variable $z$ does not appear in any $f$-term of $B_1$ as
all $f$-terms are frozen in $\exists z~B_1$.  The variables free in
$\Delta'$ are free in $\Gamma_1$, in $B_1$ or in $B_2$. Hence $z$
is not free in $\Delta'$ and from $\Gamma_1, \Delta' \vdash
\exists z~B_1$ and $\Gamma_1, \Delta', B_1 \vdash B_2$ we can
deduce $\Gamma_1, \Delta' \vdash B_2$.

From the proof of $\Gamma, \Delta' \vdash B$ we
eliminate one by one all total instances of $A$ corresponding to terms not
occurring in $\Gamma$ and $B$ using Proposition \ref{erase}, starting
from 
the largest.  Finally, we add, by weakening, the total instances 
corresponding to terms occurring in $\Gamma$ and $B$ that would not be
in $\Delta'$ and we get this way a proof of $\Gamma, \Delta \vdash
B$. 
\end{itemize}}

\medskip
\noindent
{\em Remark.} If the proposition $\forall \overline{x} \exists y~A$ is
an axiom of $\Gamma$, then we obtain a conservative extension of
$\Gamma$ by adding the axiom $\forall
\overline{x}~(f(\overline{x})/y)A$ to $\Gamma$.  The axiom $\forall
\overline{x} \exists y~A$, that it is a consequence of $\forall
\overline{x}~(f(\overline{x})/y)A$, is then redundant and can be
dropped.  Thus, we obtain also a conservative extension if we replace
the axiom $\forall \overline{x} \exists y~A$ by $\forall
\overline{x}~(f(\overline{x})/y)A$.

\section*{Acknowledgements}

The authors want to thank Jeremy Avigad, Marc Bezem, Thierry Coquand,
Dimitri Hendriks, Olivier Hermant and Helmut Schwichtenberg for
several discussions on Skolem theorem.


\begin{thebibliography}{99}
\bibitem{Avigad} J.~Avigad, {\em Eliminating definitions and skolem
functions in first-order logic}, ACM Transactions on Computational
Logic, 4, 3, 2003, pp. 402-415. 

\bibitem{Maehara} S.~Maehara, {\em The predicate calculus with
$\varepsilon$-symbol}, Journal of the Mathematical Society of Japan,
7, 4, 1955, p. 323-344.

\bibitem{Schwichtenberg} H.~Schwichtenberg, {\em 
Logic and the axiom of choice}, Logic Colloquium 78, M.~Boffa, D.~van
  Dalen, and K.~McAloon (eds.), North-Holland, 1979, pp. 351-356.

\bibitem{Mints} G.~Mints, {\em Axiomatization of a Skolem function in 
intuitionistic logic}, M.~Faller, S.~Kaufmann, and M.~Pauly (eds.)
Formalizing the Dynamics of Information, CSLI Publications, 2000,
pp. 195-114. 

\bibitem{Prawitz} D. Prawitz, {\em Natural deduction}, Amlqvist \&
Wiksell, 1965.
\end{thebibliography}
\end{document}